\documentclass[11pt,a4paper]{article}

\usepackage{jheppub}
\usepackage{amsmath}
\usepackage{amssymb}
\usepackage{amsthm}
\usepackage{braket}
\usepackage{bbold}
\usepackage{enumerate}
\usepackage{makecell}
\usepackage{mathtools}
\usepackage{hyperref}
\usepackage{tabu}
\usepackage{tikz}
\usepackage{extpfeil}
\usepackage{tabularx}
\usepackage{graphicx}
\usepackage{color}
\usepackage{lscape}
\usetikzlibrary{decorations.pathmorphing,cd}
\usepackage{pdflscape}
\usepackage{comment}
\usepackage[utf8]{inputenc}

	\usepackage{enumitem}

\newcommand{\bZ}{\mathbb{Z}}

\newcommand{\bC}{\mathbb{C}}
\newcommand{\bN}{\mathbb{N}}

\newcommand{\cN}{\mathcal{N}}

\newcommand{\Hom}{\text{Hom}}
\renewcommand{\d}{\text{d}}
\newcommand{\id}{\text{id}}

\newcommand\End{\text{End}}

\newcommand\cone{\text{cone}}

\newcommand\obj{\text{obj}}
\newcommand\bP{{\bar P}}

\newcommand{\aeq}[1]{\begin{equation}\begin{aligned}#1\end{aligned}\end{equation}}

\newcommand{\uaeq}[1]{\begin{equation*}\begin{aligned}#1\end{aligned}\end{equation*}}

\newcommand{\beq}{\begin{equation}}
\newcommand{\eeq}{\end{equation}}
\newcommand{\baeq}{\begin{equation}\begin{aligned}}
\newcommand{\eaeq}{\end{aligned}\end{equation}}

\usetikzlibrary{decorations.markings}
	\definecolor{WScolor}{RGB}{191,191,255}
	\definecolor{WScolor light}{RGB}{224,224,255}
	\definecolor{DefectColor}{RGB}{0,0,128}
	\definecolor{SymmetryDefectColor}{RGB}{0,127,0}

\tikzset{snake/.style={decorate, decoration=snake}}
\tikzset{
	symmetry defect/.style={color=SymmetryDefectColor, line width=1.5}, 
	defect/.style={color=DefectColor, line width=1.5}, 
	arrow position/.style={postaction={decorate,decoration={
		markings,
		mark=at position #1 with {\arrow{>}}
	}}},
	opp arrow position/.style={postaction={decorate,decoration={
		markings,
		mark=at position #1 with {\arrow{<}}
	}}},
	defect node/.style={circle,inner sep=1.5pt,fill, DefectColor}, 
	symmetry node/.style={circle,inner sep=1.5pt, fill, color=SymmetryDefectColor},
	defect unit node/.style={draw=DefectColor, circle, inner sep=0, minimum width=3, line width=1.3},
	unit node/.style={draw=SymmetryDefectColor, circle, inner sep=0, minimum width=3, line width=1.3}, 
	dot/.style={circle,inner sep=1pt,fill},
	black node/.style={circle,inner sep=1.5pt,fill},
	number node/.style={draw=SymmetryDefectColor, circle,inner sep=1.5pt, DefectColor} 
}

\usetikzlibrary{patterns,snakes,arrows}

\usetikzlibrary{decorations.pathmorphing,shapes}

\newcounter{sarrow}
\newcommand\xrsquigarrow[1]{%
\stepcounter{sarrow}
\begin{tikzpicture}[decoration=snake]
\node (\thesarrow) {\strut#1};
\draw[->,decorate] (\thesarrow.south west) -- (\thesarrow.south east);
\end{tikzpicture}%
}

\tikzset{
	slim defect/.style={color=DefectColor, line width=1}
}
\definecolor{boundaryColor}{RGB}{90,60,8}

\title{Complementary Projection Defects and Decomposition}
\author[1]{Fabian Klos,}
\author[2]{Daniel Roggenkamp}
\affiliation[1]{Institut f\"ur Theoretische Physik, Universit\"at Heidelberg,\\
Philosophenweg 19, 69120 Heidelberg, Germany}
\affiliation[2]{Institut f\"ur Mathematik, Universit\"at Mannheim,\\
B6, 26,  68131 Mannheim, Germany}

\abstract{
As put forward in \cite{Klos:2019axh} topological quantum field theories can be projected using so-called projection defects. The projected theory and its correlation functions can be completely realized within the unprojected one. 
An interesting example is the case of topological quantum field theories associated to IR fixed points of renormalization group flows, which 
by this method can be realized inside the theories associated to the UV. \\
In this note we show that projection defects in triangulated defect categories (such as defects in 2d topologically twisted $\cN=(2,2)$ theories) always come with complementary projection defects, and that the unprojected theory
decomposes into the theories associated to the two projection defects.
We demonstrate this in the context of Landau-Ginzburg orbifold theories.}
 
\begin{document}
\maketitle

\section{Introduction}
\label{sec:theory}

In \cite{Klos:2019axh} it was shown how two-dimensional topological quantum field theories can be projected using so-called projection defects. All objects associated to the new projected theory, such as correlation functions, categories of boundary conditions, etc.~can be explicitly described within the unprojected theory. An interesting example of this are topological quantum field theories related by a renormalization group flow\footnote{For instance as topological twists of the UV and IR fixed points of an RG flow of $\cN=(2,2)$ supersymmetric quantum field theories.}, where the theory associated to the IR fixed point can be obtained as a projection of the theory in the UV.

The defects by which such projections can be implemented are defect lines with special properties. They are idempotent with respect to fusion\footnote{Two parallel defects fuse to a new defect when brought on top of each other.}:
\[\tikz[baseline=10]{
	\fill[WScolor] (0,0) rectangle (3,1);
	\draw[defect] (1,0) -- (1,1);
	\draw[defect] (2,0) -- (2,1);
	\node at (1.5,-.2) {$P\otimes P$};
}\cong\tikz[baseline=10]{
	\fill[WScolor] (0,0) rectangle (3,1);
	\draw[defect] (1.5,0) -- (1.5,1);
	\node at (1.5,-.2) {$P$};.
}\]
In addition they have to have a counit or unit. A counit of $P$ is a morphism (defect changing field) 
$c:P\rightarrow I$ from $P$ to the 
identity (or invisible) defect $I$ of the theory which satisfies 
	\aeq{\label{eq:counit conditions}\tikz[baseline=20]{
		\fill[WScolor] (0,0) rectangle (1.5,1.5);
		\draw[defect] (1,0) -- (1,1.5);
		\draw[densely dashed] (.5,1.1) -- (.5,1.5);
		\draw[defect] (1,.25) .. controls (.5,.5) .. (.5,1);
		\node[unit node, DefectColor] at (.5,1.05) {};
		\node at (1.2,.2) {$P$};
		\node at (.3,1.05) {$c$};
		
	}=\tikz[baseline=20]{
		\fill[WScolor] (0,0) rectangle (1.5,1.5);
		\draw[defect] (1,0) -- (1,1.5);
		\draw[densely dashed] (1,.75) .. controls (.5,1) .. (.5,1.5);
		\node at (1.2,.2) {$P$};
	}\qquad\text{and}\qquad\tikz[baseline=20,xscale=-1]{
		\fill[WScolor] (0,0) rectangle (1.5,1.5);
		\draw[defect] (1,0) -- (1,1.5);
		\draw[densely dashed] (.5,1.1) -- (.5,1.5);
		\draw[defect] (1,.25) .. controls (.5,.5) .. (.5,1);
		\node[unit node, DefectColor] at (.5,1.05) {};
		\node at (1.2,.2) {$P$};
		\node at (.3,1.05) {$c$};
	}=\tikz[baseline=20,xscale=-1]{
		\fill[WScolor] (0,0) rectangle (1.5,1.5);
		\draw[defect] (1,0) -- (1,1.5);
		\draw[densely dashed] (1,.75) .. controls (.5,1) .. (.5,1.5);
		\node at (1.2,.2) {$P$};
	}}
In these diagrams, which we read from bottom to top, dashed lines always represent the identity defect. The identity defect exists in any theory and does not affect correlation functions when it is inserted. Moreover, there are 
natural junction fields by which it can end (respectively begin) on any other defect. The trivalent junctions of $P$ are given by the isomorphisms $P\otimes P\cong P$. 

A unit of $P$ is a morphism $u:I\rightarrow P$ from the identity defect to $P$, satisfying the 
corresponding relations obtained by vertical reflection of the diagrams above.

Given such a projection defect $P$, the $P$-projected theory
can  be described in terms of the unprojected theory as follows. While bulk fields of the  unprojected theory can be regarded as endomorphisms (i.e.~defect fields) of the identity defect $I$, bulk fields of the projected theory can be realized as endomorphisms of $P$ in the unprojected theory.
Boundary conditions of the projected theory can be represented as boundary conditions $B$ of the unprojected theory, which are invariant under fusion with $P$, i.e.~$B\cong P\otimes B$, and defect lines in the projected theory correspond to defect lines $D$ in the unprojected theory, which are invariant under fusion with $P$ from both sides, $D\otimes P\cong D\cong P\otimes D$. 

Indeed, to every object (such as fields, boundary conditions, defects etc.) in the unprojected theory, one can  associate a respective object of the projected theory by surrounding it with the defect $P$. For instance,
bulk fields $\phi$ of the unprojected theory can be mapped to bulk fields of the projected theory by encircling their insertions with $P$, and fusion with $P$ maps any boundary condition $B$ of the unprojected theory to a $P$-invariant one
representing a boundary condition of the projected theory:
\begin{equation}\label{eq:P projected}\tikz[baseline=25]{
	\fill[WScolor] (0,0) rectangle (2,2);
	\draw[densely dashed] (1,.5) .. controls (.5,.5) and (.5,1.5) .. (1,1.5);
	\draw[defect] (1,0) -- (1,2);
	\node at (1.3,.2) {$P$};
	\node[defect node] at (.63,1) {};
	\node at (.4,1) {$\phi$};
}=\tikz[baseline=25]{
	\fill[WScolor] (0,0) rectangle (2,2);
	\draw[defect] (1,0) -- (1,.5);
	\draw[defect] (1,1.5) -- (1,2);
	\draw[defect] (1,1) circle (.5);
	\node at (1.3,.2) {$P$};
	\node[defect node] at (1.2,1) {};
	\draw[densely dashed] (1.2,.6) -- (1.2,1.4);
	\node at (1,1) {$\phi$};
}=\tikz[baseline=25]{
	\fill[WScolor] (0,0) rectangle (2,2);
	\draw[densely dashed] (1,.5) .. controls (1.5,.5) and (1.5,1.5) .. (1,1.5);
	\draw[defect] (1,0) -- (1,2);
	\node at (.7,.2) {$P$};
	\node[defect node] at (1.37,1) {};
	\node at (1.6,1) {$\phi$};
}\qquad\qquad\tikz[baseline=25]{
	\fill[WScolor] (0,0) rectangle (2,2);
	\draw[defect, boundaryColor] (2,0) -- (2,2);
	\draw[defect] (1,0) -- (1,2);
	\node at (.8,.2) {$P$};
	\node at (2.2,.2) {$B$};
}\end{equation}

Correlation functions of the projected theory can then be obtained as correlation functions of the unprojected theory with a network of the defect $P$ inserted. 
Indeed, counitality (unitality) of $P$  ensures that $P$ splits as $P\cong {}^\dagger R\otimes R$ ($P\cong {
R}^\dagger \otimes R$), 
where $R$ is a defect between the unprojected and projected theory 
and ${}^\dagger R$ ($R^\dagger$) is an adjoint, i.e.~an opposite oriented version of it.\footnote{Left adjoints ${}^\dagger R$ and right adjoints $R^\dagger$ of a defect $R$ can be thought of as left-, respectively right-bended versions of the defect $R$.} Moreover, $R$ has invertible right quantum dimension:
\[\tikz[baseline=25]{
	\fill[WScolor light] (-1,0) rectangle (3,2);
	\fill[WScolor] (1.5,1) ellipse (1 and .6);
	\draw[densely dashed] (1.5,1.6) -- (1.5,2);
	\draw[densely dashed] (1.5,.4) -- (1.5,0);
	\draw[defect,opp arrow position=.5] (1.5,1) ellipse (1 and .6);
	\node at (1.5,1) {$\substack{\text{unproj.}\\ \text{TQFT}}$};
	\node at (-.3,.5) {$\substack{\text{proj.}\\ \text{TQFT}}$};
	\node at (.3,1.4) {$R$};
}=\tikz[baseline=25]{
	\fill[WScolor light] (0,0) rectangle (2,2);
	\draw[densely dashed] (1,0) -- (1,2);
}\]
This allows to apply the trick underlying the generalized orbifold construction \cite{Frohlich:2009gb}.
Inserting bubbles of the unprojected theory into correlation functions of  the projected theory and then expanding the bubbles until they cover all of space-time one recovers correlation functions of the projected theory as correlation functions of the unprojected theory with networks of the defect $P$ inserted.
\[\tikz[baseline=-15]{
	\fill[WScolor light] (0,0) arc (180:0:.6) arc (180:360:.3) arc (180:0:.3) arc (360:180:1.2);
	\draw[densely dashed] (2,-.9) -- (1.5,-.3);
	\node[defect node] at (1.75,-.6) {};
	\draw[defect, opp arrow position=.6, color=boundaryColor] (0,0) arc (180:0:.6) arc (180:360:.3) arc (180:0:.3) arc (360:180:1.2);
	\node[defect node, color=boundaryColor] at (.17,-.6) {};
	\node at (2.5,-.5) {$B$};
	\node at (.7,-.3) {$\substack{\text{proj.}\\ \text{TQFT}}$};
}\quad=\quad\tikz[baseline=-15]{
	\fill[WScolor light] (0,0) arc (180:0:.6) arc (180:360:.3) arc (180:0:.3) arc (360:180:1.2);
	\draw[densely dashed] (2,-.9) -- (1.5,-.3);
	\node[defect node] at (1.75,-.6) {};
	\draw[defect, opp arrow position=.6, color=boundaryColor] (0,0) arc (180:0:.6) arc (180:360:.3) arc (180:0:.3) arc (360:180:1.2);
	\node[defect node, color=boundaryColor] at (.17,-.6) {};
	\node at (2.5,-.5) {$B$};
	\fill[WScolor] (.7,-.3) ellipse (.3 and .5);
	\draw[defect,opp arrow position=.5] (.7,-.3) ellipse (.3 and .5);
	\fill[WScolor] (2.1,-.2) circle (.2);
	\draw[defect] (2.1,-.2) circle (.2);
	\node at (.3,.2) {$R$};
}\quad=\quad\tikz[baseline=-15]{
	\fill[WScolor] (0,0) arc (180:0:.6) arc (180:360:.3) arc (180:0:.3) arc (360:180:1.2);
	\draw[defect] (2,-.9) -- (1.5,-.3);
	\node[defect node] at (1.75,-.6) {};
	\draw[defect, opp arrow position=.6, color=boundaryColor] (0,0) arc (180:0:.6) arc (180:360:.3) arc (180:0:.3) arc (360:180:1.2);
	\node[defect node, color=boundaryColor] at (.17,-.6) {};
	\node at (3,-.5) {${}^\dagger R\otimes B$};
	\node at (1.45,-.65) {$P$};
	\node at (.7,-.3) {$\substack{\text{unproj.}\\ \text{TQFT}}$};
}\]

An interesting class of examples of projection defects arises from RG flows in $\cN=(2,2)$ superconformal field theories.
As put forward in \cite{Brunner:2007ur}, applying an RG flow on only one side of the identity defect in the UV theory gives rise to a defect $R$ between the UV and the IR theory. 
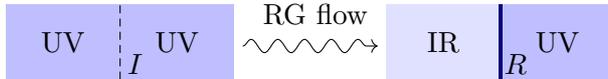
\[\tikz[baseline=10]{
	\fill[WScolor] (-.5,0) rectangle (2.5,1);
	\draw[densely dashed] (1,0) -- (1,1);
	\node at (1.2,.2) {$I$};
	\node at (.25,.5) {UV};
	\node at (1.75,.5) {UV};
}\;\xrsquigarrow{ RG flow }\;\tikz[baseline=10]{
	\fill[WScolor light] (-.5,0) rectangle (1,1);
	\fill[WScolor] (1,0) rectangle (2.5,1);
	\draw[defect] (1,0) -- (1,1);
	\node at (1.2,.2) {$R$};
	\node at (.25,.5) {IR};
	\node at (1.75,.5) {UV};
}\]
This RG defect projects UV degrees of freedom into the IR. The topologically twisted version of this defect has invertible right quantum dimension, and $P={}^\dagger R\otimes R$ is a counital\footnote{$P'=R^\dagger\otimes R$ is a unital projection defect.} projection defect realizing the (topologically twisted) IR theory inside the theory in the UV.

For more details on projections of topological quantum field theories by means of projection defects we refer the reader to \cite{Klos:2019axh}.

In this note we consider topological quantum field theories whose defect categories are tensor triangulated, such as topologically twisted $\cN=(2,2)$ superconformal field theories. We show that in this setup any counital projection defect $P$ comes with a complementary unital projection defect $\bP$, and vice-versa, and that the unprojected theory decomposes into the $P$-projected and $\bP$-projected theories. This will be spelled out in section~\ref{sec:proof}. In section~\ref{sec:LG example} we will illustrate it in the example of projection defects in B-twisted Landau-Ginzburg orbifold models.

\section{Complementary Projection Defects}
\label{sec:proof}

From now on, we assume that the defect category is tensor triangulated as explained shortly. We will show that every counital projection defect $P$ comes with a complementary unital projection defect $\bP$ and vice-versa.
Complementarity of a pair $(P,\bP)$ means that
\begin{equation}\begin{aligned}\label{eq:complementary conditions}
	&\text{ mutual fusion vanishes }P\otimes\bP\cong 0 \cong \bP\otimes P\,,\;\text{and} \\
	&\text{ the identity defect is isomorphic to a cone } I\cong\cone(s:\bP[-1]\rightarrow P)\,.
\end{aligned}\end{equation}
In fact, we can also start with two defects $P$ and $\bP$ which are idempotent with respect to fusion and satisfy conditions \eqref{eq:complementary conditions}. Then, $P$ carries a unit and $\bP$ a counit. Physically, the second condition in \eqref{eq:complementary conditions} means that the sum $P\oplus\bP$ of the two projectors can be deformed (or perturbed) to the identity defect.

\subparagraph{Triangulated defect categories.} We start by briefly collecting some basic properties of triangulated tensor categories. For the full set of axioms see e.g.~\cite{Orlov:2003yp}. A tensor triangulated category $T$
comes with automorphisms $[n]$ called shift functors for all $n\in\mathbb{Z}$. Moreover, it comes with a collection of exact (or distinguished) triangles 
\begin{equation}\label{eq:extr}
C\stackrel{\phi}{\rightarrow} D\stackrel{\psi}{\rightarrow} E\stackrel{\chi}{\rightarrow} C[1]
\end{equation}
satisfying a list of axioms. Here $C,D,E\in\text{obj}(T)$. For instance, 
for every morphism $\phi:C\rightarrow D$ between objects $C,D\in\text{obj}(T)$ there is an object $\cone(\phi)\in\text{obj}(T)$ which fits into an exact triangle
\begin{equation}\label{eq:cone1}\begin{tikzcd}
	C\arrow{r}{\phi}&D\arrow{r}&\cone(\phi) \arrow{r} & C[1].
\end{tikzcd}
\end{equation}
The cone of the identity morphism $\id_C$ of any object $C$ is trivial, $\cone(\id_D)=0$.
Moreover,
the shift functor takes exact triangles to exact triangles, and the triangle \eqref{eq:extr} is exact if and only if the  rotated triangle
\begin{equation*}
D\stackrel{\psi}{\rightarrow} E\stackrel{\chi}{\rightarrow} C[1]\stackrel{-\phi[1]}{\rightarrow} D[1]
\end{equation*}
is exact.

A morphism of triangles consists of morphisms, $c,d,e$ such that all squares in 
\[\begin{tikzcd}
	C'\arrow{r}{\phi'}\drar[phantom, "\square"]&D'\arrow{r}{\psi'}\drar[phantom, "\square"]&E' \arrow{r}{\chi'}\drar[phantom, "\square"] & C'[1] \\
	C\arrow{r}{\phi} \arrow{u}{c}&D\arrow{r}{\psi} \arrow{u}{d}&E \arrow{r}{\chi} \arrow{u}{e} & C[1] \arrow{u}[swap]{c[1]}
\end{tikzcd}\]
commute. If $c,d,e$ are isomorphisms, they define an isomorphism of triangles. Another important property of triangulated categories is that any triangle isomorphic to an exact triangle is itself exact. Moreover, for exact triangles, 
the existence of the first two (and hence also the fourth) vertical morphisms in
\[\begin{tikzcd}
	C'\arrow{r}{\phi'}\drar[phantom, "\square"]&D'\arrow{r}{\psi'}\drar[phantom, "\square"]&E' \arrow{r}{\chi'}\drar[phantom, "\square"] & C'[1] \\
	C\arrow{r}{\phi} \arrow{u}&D\arrow{r}{\psi} \arrow{u}&E \arrow{r}{\chi} \arrow[dashed]{u} & C[1] \arrow{u}
\end{tikzcd}\]
implies the existence of the dashed morphism with all squares commuting. The morphism induced in this way by two isomorphisms is itself an isomorphism.

Since our aim is to describe defects, we require $T$ to be a tensor category, in particular
it comes with a product $\otimes$ on objects and morphisms with neutral element $I\in\obj(T)$. 
Moreover, this product is compatible with the triangulated structure. More precisely, it 
commutes with the shift functor $[n]$ and cone construction of $T$, i.e.
\begin{equation}\label{eq: triangulated conditions}\begin{aligned}
	&(D\otimes E)[1] \cong D[1]\otimes E \cong D\otimes E[1] \\
	&\cone(\phi:D\rightarrow E)\otimes F \cong \cone(\phi\otimes\id_F:D\otimes F\rightarrow E\otimes F)
\end{aligned}\end{equation}
To put it differently, 
 the induced functors $\cdot\otimes D$ and $D\otimes\cdot$ for any $D\in\obj(T)$ are triangulated, i.e.~compatible with the triangulated structure. Furthermore, morphisms are taken to commute with the isomorphisms $D\cong I\otimes D$ and $D\cong D\otimes I$. This is expressed by the existence of four commutative diagrams of the type
\[\begin{tikzcd}
	I\otimes D\arrow{r}{\id\otimes\phi}\drar[phantom, "\square"]&I\otimes E \\
	D\arrow{r}[swap]{\phi} \arrow{u}{\sim}&E \arrow{u}[swap]{\sim}
\end{tikzcd}\qquad\qquad\tikz[baseline=30]{
	\fill[WScolor] (0,0) rectangle (2,2);
	\draw[densely dashed] (1,.3) .. controls (.6,.4) .. (.5,2);
	\draw[defect] (1,0) -- (1,2);
	\node[defect node] at (1,1) {};
	\node at (1.2,.2) {$D$};
	\node at (1.3,1) {$\phi$};
	\node at (1.2,1.8) {$E$};
}=\tikz[baseline=30]{
	\fill[WScolor] (0,0) rectangle (2,2);
	\draw[densely dashed] (1,1.3) .. controls (.6,1.4) .. (.5,2);
	\draw[defect] (1,0) -- (1,2);
	\node[defect node] at (1,1) {};
	\node at (1.2,.2) {$D$};
	\node at (1.3,1) {$\phi$};
	\node at (1.2,1.8) {$E$};
}\]
On the right hand side we have provided the relation in string diagram notation.
The other three diagrams can be obtained by mirroring the given defect diagrams.

In the following we will use the triangulated structure to show our claim. We will do this in three steps. First, we will show that every counital projection defect comes with a complementary unital projection defect. Secondly, we will argue that conversely, every unital projection defect also comes with a complementary  counital projection defect. And thirdly, we will show that for any pair $(P,\bP)$ of complementary idempotent defects, $P$ is counital and $\bP$ is unital. Afterwards we will discuss how, given such a pair of complementary projection defects, the host theory decomposes into the projected theories associated to $P$ and $\bP$.

\subparagraph{(i) Counital projections have unital complementary projections.} 
Given a counital projection defect $P\in\obj(T)$. It satisfies  $P\otimes P\cong P$ and there is a morphism $c:P\rightarrow I$ such that the following two squares commute:
\[\begin{tikzcd}
	P \otimes P \arrow{r}{c\otimes\id_P}\drar[phantom, "\square"] & I \otimes P & P\otimes P \arrow{r}{\id_P\otimes c}\drar[phantom, "\square"] & P\otimes I \\
	P \arrow[shift right]{u}{\sim} \arrow{r}[swap]{\id_P} & P \arrow[shift right]{u}{\sim} & P \arrow{r}[swap]{\id_P} \arrow[shift right]{u}{\sim} & P \arrow[shift right]{u}{\sim}
\end{tikzcd}\]
These two relations are just the counit conditions  \eqref{eq:counit conditions}.
Defining $\bP:=\cone(c:P\rightarrow I)$, the exact triangle with respect to the counit becomes
\begin{equation}\label{eq:counit triangle}\begin{tikzcd}
	P\arrow{r}{\text{c}} &I \arrow{r}{u} & \bP \arrow{r}{s[1]} & P[1].
\end{tikzcd}\end{equation}
The key idea is now the following: $c$ obeys counit conditions, $u$ obeys unit conditions and $\cone(s)\cong I$  is isomorphic  to the  identity defect. Given any of the three morphisms $s$, $c$ or $u$ satisfying the respective condition, the triangulated structure implies that the other morphisms exist and satisfy the respective conditions. Let us spell this out for the case at hand. 

Applying $\cdot\otimes P$ to \eqref{eq:counit triangle} one obtains the upper row of 
\[\begin{tikzcd}
	P \otimes P \arrow{r}{c\otimes\id_P}\drar[phantom, "\square"] & I \otimes P \arrow{r} & \bP\otimes P \arrow{r} & P[1]\otimes P \\
	P \arrow[shift right]{u}{\sim} \arrow{r}{\id_P} & P \arrow[shift right]{u}{\sim} \arrow{r} & 0 \arrow{r}\arrow[shift right,dashed]{u} & P[1] \arrow[shift right]{u}{\sim}
\end{tikzcd}\]
Since  $\cdot\otimes P$ is a triangulated functor, it is an exact triangle. The counit conditions gives rise to the left commuting square. 
By the axioms of triangulated categories, the lower triangle is also exact, the dashed morphism exists and all squares commute. Also, the dashed morphism is an isomorphism. Hence, $\bP\otimes P\cong 0$. Similar considerations lead to $P\otimes \bP\cong 0$.

Next, applying the functor $\bP\otimes\cdot$ to \eqref{eq:counit triangle} yields
\[\begin{tikzcd}
	\bP \otimes P \arrow{r}{\id_{\bP}\otimes c}\drar[phantom, "\square"] & \bP \otimes I \arrow{r}{\id_{\bP}\otimes u} & \bP\otimes \bP \arrow{r} & \bP \otimes P[1]  \\
	0 \arrow[shift right]{u}{\sim} \arrow{r} & \bP \arrow[shift right]{u}{\sim} \arrow{r}{\id_\bP} & \bP \arrow{r}\arrow[shift right,dashed]{u} & 0 \arrow[shift right]{u}{\sim}
\end{tikzcd}.\]
Again, both rows are exact triangles. 
Because morphisms from the zero object are unique, the first square commutes. Hence, the dashed morphism exists, makes all squares commute and is an isomorphism. Therefore, $\bP$ is idempotent with respect to $\otimes$ and the first of the two unit conditions
\[\begin{tikzcd}
	\bP \arrow{r}{\id_{\bP}} \drar[phantom, "\square"]& \bP & \bP \arrow{r}{\id_\bP}\drar[phantom, "\square"] & \bP\\
	\bP \otimes I \arrow[shift right]{u}{\sim} \arrow{r}{\id_\bP\otimes u} & \bP\otimes\bP \arrow[shift right]{u}{\sim} & I\otimes\bP \arrow{r}{u\otimes\id_\bP} \arrow[shift right]{u}{\sim} & \bP\otimes\bP\arrow[shift right]{u}{\sim}
\end{tikzcd}\]
holds. In the same way, 
application of $\cdot\otimes \bP$ to \eqref{eq:counit triangle} implies the second unit condition.
$\bP$ is therefore a unital projection defect.

It remains to show that the  identity defect $I$ is isomorphic to a cone of  a morphism $\bP[-1]\rightarrow P$. But this follows by rotating the exact triangle
\eqref{eq:counit triangle}  to
\uaeq{
	\bP[-1] \xrightarrow{s} P\xrightarrow{{c}} I \xrightarrow{{u}} \bP
}

\subparagraph{(ii) Unital projections have counital complementary projections.} The above arguments also work the other way around: the existence of an idempotent $\bP\in\obj(T)$ with unit $u:I\rightarrow \bP$ implies the existence of a complementary counital idempotent
\[
	P:=\text{cone}(u[-1]:I[-1]\rightarrow \bP[-1])\in\obj(T)\,.
\]
Of course taking this projection $P$ and applying the construction in (i) gives back the original projection $\bP$. Namely, 
the counit of  $P$ is the left morphism in \eqref{eq:counit triangle} and automatically $\cone(c:P\rightarrow I)\cong \bP$. 

Vice-versa, starting with a unital projection defect $P$ and first constructing the counital projection defect $\bP$ as in (i) and then applying the construction in (ii) returns the original projection defect $P$.

\subparagraph{(iii) Complementary projections are (co)unital.} We now turn the above discussion around and start with two idempotents $P, \bar{P}\in\obj(T)$ satisfying
\uaeq{
P\otimes P\cong P\,,\qquad
\bP\otimes \bP\cong\bP\,,\qquad
P\otimes\bP\cong 0\cong \bP\otimes P\,.
}
Moreover, we  assume that there is  a morphism 
 $s:\bar P[-1]\rightarrow P$ such that $\cone(s)\cong I$. In other words,
\uaeq{
	\bP[-1]\xrightarrow{s} P \rightarrow I \rightarrow \bP
}
is exact. Application of $P\otimes\cdot$ to this triangle gives rise to the first row of 
\[\begin{tikzcd}
	P \otimes\bP[-1] \arrow{r}{\id_P\otimes s} \drar[phantom, "\square"]& P \otimes P \arrow{r} & P\otimes I \arrow{r} & P \otimes \bP \\
	0 \arrow[shift right]{u}{\sim} \arrow{r} & P \arrow[shift right]{u}{\sim} \arrow{r}{\id_P} & P \arrow{r}\arrow[shift right,dashed]{u}{\sim} & 0 \arrow[shift right]{u}{\sim}
\end{tikzcd}.\]
The first square commutes because there is a unique morphism from the zero object into any object. Hence, the middle square also commutes and yields the second counit condition. Similarly, application of the exact functors $\cdot\otimes P$,  $\bP\otimes\cdot$ and $\cdot\otimes\bP$ leads to the first counit condition on $P$ and the unit conditions on $\bP$. Hence, of two complementary projectors one is always unital and the other counital.

Next, we will  show how, given a  complementary pair $(P,\bP)$ of projection defects, 
the unprojected theory  decomposes into the 
$P$-projected theory and the $\bP$-projected theory. Let us start with the spectrum of boundary conditions.

\subparagraph{Decomposition -- boundary spectrum.} The category of boundary conditions of the unprojected theory decomposes  into the subcategories of boundary conditions of the two projected theories.
Namely, every boundary condition $B$ in the unprojected theory can be expressed as
\begin{equation}\label{eq:boundary composition}\begin{aligned}
	B\cong I\otimes B &\cong \cone(s:\bP[-1]\rightarrow P) \otimes B \\
	&\cong \cone(s\otimes\id_B:\bP \otimes B[-1]\rightarrow P \otimes B)
\end{aligned}\end{equation}
Hence, every boundary condition in the  unprojected theory  is a cone of a morphism $s\otimes\id_B$ from a $\bP$-invariant boundary condition to a $P$-invariant boundary condition. 
The category of boundary conditions in the unprojected theory is therefore generated by the subcategories of $\bP$- and $P$-invariant boundary conditions.
The latter correspond to the categories of boundary conditions in the $\bP$- and $P$ projected theories, respectively. By complementarity, the two subcategories are disjoint
\[\begin{aligned}
	P\otimes B \cong B \quad&\Rightarrow \quad \bP \otimes B\cong\bP\otimes P\otimes B\cong 0 \\
	\bP\otimes B \cong B \quad&\Rightarrow \quad P \otimes B\cong P\otimes\bP\otimes B\cong 0 \\
\end{aligned}\]
and due to  \eqref{eq:boundary composition} all boundary conditions in the kernel of $P\otimes\cdot$ are $\bP$-invariant and vice-versa:
\[\begin{aligned}
	P\otimes B \cong 0 \quad&\Rightarrow \quad \bP \otimes B\cong B \\
	\bP\otimes B \cong 0 \quad&\Rightarrow \quad P \otimes B\cong  B \\
\end{aligned}\]

\subparagraph{Decomposition -- bulk spectrum.}
Similarly, the bulk spectrum of the unprojected theory can be reconstructed from the bulk spectra of the two projected theories, once $s$ is known.

First, every bulk field $\phi:I\rightarrow I$ induces an endomorphism  of $P$ by enclosing it with the appropriate projection defect.\footnote{There is also an equivalent point of view: Concatenation with the counit (unit) gives a morphism $P\xrightarrow{\phi\circ c}I$ ($I\xrightarrow{u\circ\phi}\bP$) and the morphisms $P\rightarrow I$ ($I\rightarrow \bP$) can be seen to be one-to-one with the endomorphisms of $P$ ($\bP$). In fact, one can regard the (co)unit as the one-dimensional equivalent of an RG defect mapping bulk fields of the unprojected (UV) into the projected theory (IR). This connects to ideas in \cite{Konechny:2012wm} and might be interesting for dimensions greater than two \cite{Carqueville:2017aoe} where the RG defects of this note appear as a form of (co)unit.} Because of the counit and projection properties of $P$, this endomorphism, which we call $\alpha(\phi)$ can be written in several ways, see \eqref{eq:P projected}. The first one is 
\begin{equation}\label{eq:bulk enclosed}
\begin{tikzcd}
	\alpha(\phi):\;P\arrow{r}{\sim} & I\otimes P\arrow{r}{\phi\otimes\id_P} & I\otimes P\arrow{r}{\sim} & P.
\end{tikzcd}\end{equation}
The same holds for the endomorphisms $\bar{\alpha}(\phi)$ induced on the complementary unital projector  $\bP$:
\begin{equation}
\begin{tikzcd}
	\bar{\alpha}(\phi):\;\bP\arrow{r}{\sim} & I\otimes \bP\arrow{r}{\phi\otimes\id_\bP} & I\otimes \bP\arrow{r}{\sim} & \bP.
\end{tikzcd}\end{equation}
Thus, we have a map $\End(I)\rightarrow\End(P)\oplus\End(\bar{P})$, $\phi\mapsto(\alpha(\phi),\bar{\alpha}(\phi))$. In fact, the image of this map is not $\End(P)\oplus\End(\bar{P})$ but rather $\End(\bP[-1]\stackrel{s}{\rightarrow}P)$, the pairs of morphisms $(\alpha,\bar{\alpha})\in\End(P)\oplus\End(\bar{P})$ which  are compatible with $s$, i.e.~all those $(\alpha,\bar\alpha)$ such that the following diagram commutes
\begin{equation*}
\begin{tikzcd}
	\bP[-1] \arrow{r}{s}\drar[phantom, "\square"] & P  \\
	\bP[-1] \arrow{r}{s} \arrow{u}{\bar{\alpha}[-1]} & P  \arrow{u}{\alpha} 
\end{tikzcd}
\end{equation*}
This can be read off from the first two columns of the following diagram
\[\begin{tikzcd}
	\bP[-1] \arrow{r}{s}\drar[phantom, "\square"] & P \arrow{r}{c} & I \arrow{r}{u} & \bP \\
	I\otimes \bP[-1] \arrow{r}{\id_I\otimes s} \arrow{u}{\sim}\drar[phantom, "\square"] & I\otimes P \arrow{r}{\id_I\otimes c} \arrow{u}{\sim} & I\otimes I \arrow{r}{\id_I\otimes u} \arrow[dashed]{u}{\sim} & I\otimes \bP \arrow{u}{\sim} \\
	I\otimes \bP[-1] \arrow{r}{\id_I\otimes s} \arrow{u}{\phi\otimes\id[-1]}\drar[phantom, "\square"] & I\otimes P \arrow{r}{\id_I\otimes c} \arrow{u}{\phi\otimes\id} & I\otimes I \arrow{r}{\id_I\otimes u} \arrow[dashed]{u}{\phi\otimes\id} & I\otimes \bP \arrow{u}{\phi\otimes\id} \\
	\bP[-1] \arrow{r}{s} \arrow{u}{\sim} & P \arrow{r}{c} \arrow{u}{\sim} & I \arrow{r}{u} \arrow[dashed]{u}{\sim} & \bP \arrow{u}{\sim}
\end{tikzcd}\]
Note that all squares, and in particular the ones on the left commute. Composing the vertical maps, one arrives at the following diagram.
\begin{equation}\label{eq:induced UV bulk}\begin{tikzcd}
	\bP[-1] \arrow{r}{s}\drar[phantom, "\square"] & P \arrow{r}{c} & I \arrow{r}{u} & \bP \\
	\bP[-1] \arrow{r}{s} \arrow{u}{\bar{\alpha}[-1]} & P \arrow{r}{c} \arrow{u}{\alpha} & I \arrow{r}{u} \arrow[dashed]{u}{\phi} & \bP \arrow{u}{\bar{\alpha}}
\end{tikzcd}\end{equation}
where $\alpha=\alpha(\phi)$ and $\bar{\alpha}=\bar{\alpha}(\phi)$ and 
of course all squares commute. This shows the claim that $(\alpha(\phi),\bar{\alpha}(\phi))\in\End(\bP[-1]\stackrel{s}{\rightarrow}P)$. 

On the other hand, since the rows in \eqref{eq:induced UV bulk} are exact triangles, any endomorphism $(\alpha,\bar{\alpha})\in\End(\bP[-1]\stackrel{s}{\rightarrow}P)$ gives  rise to an endomorphism  $\phi\in\End(I)$.  The latter  in turn satisfies $\alpha(\phi)=\alpha$ and $\bar{\alpha}(\phi)=\bar{\alpha}$ which follows from the commutativity of squares in \eqref{eq:induced UV bulk}, namely
\[\tikz[baseline=18]{
	\fill[WScolor] (0,0) rectangle (1.5,1.5);
	\draw[defect] (1,0) -- (1,1.5);
	\draw[densely dashed] (1,.25) .. controls (.6,.35) .. (.5,.75);
	\draw[densely dashed] (.5,.75) .. controls (.6,1.15) .. (1,1.25);
	\node[defect node] at (.5,.75) {};
	\node at (.25,.75) {$\phi$};
}=\tikz[baseline=18]{
	\fill[WScolor] (0,0) rectangle (1.5,1.5);
	\draw[defect] (1,0) -- (1,1.5);
	\draw[defect] (1,.25) .. controls (.6,.35) .. (.5,.7);
	\draw[densely dashed] (.5,.75) .. controls (.6,1.15) .. (1,1.25);
	\node[unit node,DefectColor] at (.5,.72) {};
	\node at (.25,.75) {$c$};
	\node[defect node] at (.55,1) {};
	\node at (.4,1.2) {$\phi$};
}=\tikz[baseline=18]{
	\fill[WScolor] (0,0) rectangle (1.5,1.5);
	\draw[defect] (1,0) -- (1,1.5);
	\draw[defect] (1,.25) .. controls (.6,.35) .. (.5,.7);
	\draw[densely dashed] (.5,.75) .. controls (.6,1.15) .. (1,1.25);
	\node[unit node,DefectColor] at (.5,.72) {};
	\node at (.25,.75) {$c$};
	\node[defect node] at (.6,.4) {};
	\node at (.4,.3) {$\alpha$};
}=\tikz[baseline=18]{
	\fill[WScolor] (.5,0) rectangle (1.5,1.5);
	\draw[defect] (1,0) -- (1,1.5);
	\node[defect node] at (1,.75) {};
	\node at (.75,.75) {$\alpha$};
}\]
for the counital projection defect $P$ and
\[\tikz[baseline=18]{
	\fill[WScolor] (0,0) rectangle (1.5,1.5);
	\draw[defect] (1,0) -- (1,1.5);
	\draw[densely dashed] (1,.25) .. controls (.6,.35) .. (.5,.75);
	\draw[densely dashed] (.5,.75) .. controls (.6,1.15) .. (1,1.25);
	\node[defect node] at (.5,.75) {};
	\node at (.25,.75) {$\phi$};
}=\tikz[baseline=18]{
	\fill[WScolor] (0,0) rectangle (1.5,1.5);
	\draw[defect] (1,0) -- (1,1.5);
	\draw[densely dashed] (1,.25) .. controls (.6,.35) .. (.5,.7);
	\draw[defect] (.5,.75) .. controls (.6,1.15) .. (1,1.25);
	\node[unit node,DefectColor] at (.5,.72) {};
	\node at (.25,.75) {$u$};
	\node[defect node] at (.6,.4) {};
	\node at (.4,.3) {$\phi$};
}=\tikz[baseline=18]{
	\fill[WScolor] (0,0) rectangle (1.5,1.5);
	\draw[defect] (1,0) -- (1,1.5);
	\draw[densely dashed] (1,.25) .. controls (.6,.35) .. (.5,.7);
	\draw[defect] (.5,.75) .. controls (.6,1.15) .. (1,1.25);
	\node[unit node,DefectColor] at (.5,.72) {};
	\node at (.25,.75) {$u$};
	\node[defect node] at (.55,1) {};
	\node at (.4,1.2) {$\bar\alpha$};
}=\tikz[baseline=18]{
	\fill[WScolor] (.5,0) rectangle (1.5,1.5);
	\draw[defect] (1,0) -- (1,1.5);
	\node[defect node] at (1,.75) {};
	\node at (.75,.75) {$\bar\alpha$};
}\]
for the unital projection defect $\bar P$.

Hence, $\End(I)\cong\End(\bP[-1]\stackrel{s}{\rightarrow}P)$, so the  algebra of bulk fields of the unprojected theory  can be  reconstructed from the  ones  of  the two projected theories.
This discussion naturally extends to the  fermionic bulk spectrum $\Hom(I[-1],I)$.

\section{Application to Landau-Ginzburg models}
\label{sec:LG example}

In this  section we illustrate  the construction of complementary projection defects in the context of 
 B-twisted Landau-Ginzburg models and their orbifolds. 
We start out by reviewing how defects in this setup can be described by means of matrix factorizations  \cite{Brunner:2007qu}. We will 
briefly discuss the triangulated structure of the category  of matrix factorizations  \cite{Orlov:2003yp} and then 
apply our construction to explicit projection defects in  Landau-Ginzburg orbifolds.

\subparagraph{B-type defects and matrix factorizations.} B-twisted two-dimensional $\cN=(2,2)$ Landau-Ginzburg  models are 
characterized by their chiral fields  $X_1, ..., X_N$ and a superpotential $W\in\bC[X_i]$. B-type defects $D$
between two such  models 
\uaeq{\tikz{
	\fill[WScolor] (-3,0) rectangle (3,1);
	\draw[defect] (0,0) -- (0,1);
	\node at (.2,.2) {$D$};
	\node at (-1.5,.5) {$W'(Z_i)$};
	\node at (1.5,.5) {$W(X_i)$};
}}
can  be described by matrix factorizations of the difference $W'-W$ of the superpotentials on both sides of $D$. A matrix factorization of a polynomial $W\in S:=\bC[X_i]$ consists of two free $S$-modules $D_0$ and $D_1$ of the same rank together with $S$-module homomorphisms
\uaeq{
	D: D_1\;
		\tikz[baseline=0]{
			\node at (0,.6) {$\d_1$};
			\draw[arrow position = 1] (-.5,.2) -- (.5,.2);
			\draw[arrow position = 1] (.5,0) -- (-.5,0);
			\node at (0,-.4) {$\d_0$};
		}\;
	D_0\quad\text{also written}\quad
	\d_D=\begin{pmatrix}0&\d_1\\\d_0&0\end{pmatrix}
}
which satisfy $\d_0\circ\d_1=W\cdot\id_{D1}$ and $\d_1\circ \d_0=W\cdot\id_{D0}$.

Finite-rank matrix factorizations of a fixed polynomial $W$ form a category, $\text{hmf}(W)$. Morphisms 
$D\rightarrow D'$ in this  category are `even' $S$-module homomorphisms
\[
	\phi= \begin{pmatrix}\phi_0&0\\0&\phi_1\end{pmatrix}: D_0\oplus D_1\rightarrow D_0'\oplus D_1'
\]
for which $\d_{D'}\circ\phi-\phi\circ\d_D=0$, modulo homotopy. Two even morphisms $\phi,\phi':D\rightarrow D'$ are homotopic if they differ by $\d_{D'}\circ\psi+\psi\circ\d_D$ for an `odd' $S$-module homomorphism
\[
	\psi= \begin{pmatrix}0&\psi_1\\\psi_0&0\end{pmatrix}: D_0\oplus D_1\rightarrow D_0'\oplus D_1'
\]
These morphisms correspond to defect changing fields. 

Parallel defect lines, when brought close together, fuse to new defect lines:
\uaeq{\tikz[scale=0.8,baseline=8]{
	\fill[WScolor] (-6,-0.5) rectangle (3,1.5);
	\draw[defect] (0,-.5) -- (0,1.5);
	\draw[defect] (-3,-.5) -- (-3,1.5);
	\node at (-2.7,-.1) {$D'$};
	\node at (.35,-.1) {$D$};
	\node at (-4.5,.5) {$W''(Z_i)$};
	\node at (-1.5,.5) {$W'(Y_i)$};
	\node at (1.5,.5) {$W(X_i)$};
}
\rightsquigarrow
\tikz[scale=0.8,baseline=8]{
	\fill[WScolor] (-6,-0.5) rectangle (3,1.5);
	\draw[defect] (-1.5,-.5) -- (-1.5,1.5);
		\node at (-.7,-.1) {$D'\otimes D$};
	\node at (-3.8,.5) {$W''(Z_i)$};
	\node at (1.0,.5) {$W(X_i)$};
	}
}
In the case of B-type defects in Landau-Ginzburg models, this fusion is described by the tensor product  of matrix factorizations
\uaeq{
	D\otimes D': \begin{matrix}&D_0\otimes D'_1 \\ \oplus &D_1\otimes D'_0\end{matrix}\;
		\tikz[baseline=0]{
			\node at (0,.9) {$\begin{pmatrix}-\id_{D_0}\otimes\d_1' & \d_1\otimes \id_{D_0'} \\ \d_{0}\otimes \id_{D_1'} & \id_{D_1}\otimes \d_{0}'\end{pmatrix}$};
			\draw[arrow position = 1] (-2.5,.2) -- (2.5,.2);
			\draw[arrow position = 1] (2.5,0) -- (-2.5,0);
			\node at (0,-.8) {$\begin{pmatrix}-\id_{D_0}\otimes\d_0' & \d_1\otimes \id_{D_1'} \\ \d_{0}\otimes \id_{D_0'} & \id_{D_1}\otimes \d_{1}'\end{pmatrix}$};
		}\;\begin{matrix}&D_0\otimes D'_0 \\ \oplus &D_1\otimes D'_1\end{matrix}.
}
Here, $\otimes$ is shorthand for $\otimes_{\bC[Y_i]}$. The resulting matrix factorization has to be regarded as a matrix factorization over the ring $\bC[X_i,Z_i]$ and is a priori of infinite rank. It is however isomorphic to a
 finite-rank matrix factorization \cite{Brunner:2007qu}.

Defects between Landau-Ginzburg models give rise to the structure of a $2$-category. Objects are the Landau-Ginzburg models (potentials), $1$-morphisms are the defects between different Landau-Ginzburg models (matrix factorizations of the difference of potentials) and $2$-morphisms are the defect changing fields (morphisms between matrix factorizations).

In the case at hand, the defect categories carry even more structure -- they are triangulated. The  shift of a matrix factorization $D$ is defined by
\uaeq{
	D[n]: D_{1+n\,\text{mod}\,2}\;
		\tikz[baseline=0]{
			\node at (0,.6) {$(-1)^n\d_{1+n\,\text{mod}\,2}$};
			\draw[arrow position = 1] (-1.3,.2) -- (1.3,.2);
			\draw[arrow position = 1] (1.3,0) -- (-1.3,0);
			\node at (0,-.4) {$(-1)^n\d_{n\,\text{mod}\,2}$};
		}\;
	D_{n\,\text{mod}\,2}
	}
The exact triangles arise from the cone construction. The cone of a  
 morphism $\phi:D\rightarrow E$ between two matrix factorizations $D,E\in\text{hmf}(W)$ is given by
\aeq{\label{eq:LG cones}
	\text{cone}(\phi:D\rightarrow E): E_1\oplus D_0\;
		\tikz[baseline=0]{
			\node at (0,1) {$\begin{pmatrix}\d_{E1}& \phi_0\\ &-\d_{D0}\end{pmatrix}$};
			\draw[arrow position = 1] (-1,.2) -- (1,.2);
			\draw[arrow position = 1] (1,0) -- (-1,0);
			\node at (0,-.8) {$\begin{pmatrix}\d_{E0}&\phi_1\\ &-\d_{D1}\end{pmatrix}$};
		}\;
	E_0 \oplus D_1
}
Physically this corresponds to a deformation (or perturbation) of the sum  $E\oplus D[-1]$. 
The exact triangles are those isomorphic to triangles of the form \cite{Orlov:2003yp}
\uaeq{
	D\xrightarrow{\phi} E \xrightarrow{\begin{pmatrix}\id_E\\ 0\end{pmatrix}} \cone(\phi:D\rightarrow E) \xrightarrow{\begin{pmatrix}0,-\id_{D[1]}\end{pmatrix}} D[1].
}
A straightforward computation reveals that fusion $\otimes$, shift $[\cdot]$ and cone construction all pairwise commute. In other words, shift and fusion take exact triangles to exact triangles. Hence, the defect category in Landau-Ginzburg models satisfies all the requirements of a tensor triangulated category. Thus, the construction of
complementary projection defects outlined in section~\ref{sec:proof} can be applied.

\subparagraph{Orbifolds and equivariant matrix factorizations.} The treat\-ment  of B-type defects carries over in a straightforward way to Landau-Ginzburg orbifold theories. The latter are specified  not only by chiral fields and a superpotential but in addition by an orbifold group  acting on  the chiral fields such that the superpotential is invariant.

B-type defects in such theories are again described by matrix factorizations of the difference of superpotentials, which however have to be equivariant with respect to the action  of the orbifold groups on each side of the defect. Defect changing fields correspond to morphisms of matrix factorization, which are invariant under the induced group action, and fusion corresponds to the part of the tensor product matrix factorization which is invariant under the orbifold group associated to the model squeezed in between the two defects. In  the  case of abelian  orbifold groups, the treatment of the orbifold simplifies, and defects can be described by graded matrix factorizations. For more details on defects in Landau-Ginzburg orbifolds see \cite{Brunner:2007ur}. Importantly, categories of equivariant matrix factorizations inherit the tensor triangulated structure from the categories of matrix factorizations, and hence, the construction of complementary projection defects also applies to Landau-Ginzburg orbifold theories. In the following we will outline the construction of complementary projection defects in Landau-Ginzburg orbifold theories in a concrete example.

\subparagraph{Landau-Ginzburg orbifolds $X^d/\bZ_d$.} 
We consider the Landau-Ginzburg orbifold with a single chiral field $X$, superpotential $W=X^d$, for some $d\in\mathbb{N}_{\geq 2}$, orbifolded by the group $\bZ_d$. An element $[n]\in\bZ_d$ acts on $X$ by $X\mapsto\exp(2\pi i\frac{n}{d})X$, and hence leaves the superpotential  $W$ invariant.

For $d>2$, these models exhibit relevant perturbations  by (a,c)-fields  which trigger renormalization group  flows to  Landau-Ginzburg orbifolds $X^{d'}/\bZ_{d'}$ of the same type but with $d'<d$.\footnote{These are mirror to flows between the unorbifolded Landau-Ginzburg models with superpotentials $W=X^d$ and $W'=X^{d'}$ triggered by deformation of the superpotential $W$ by lower degree polynomials.} RG defects  describing these  flows were constructed in 
\cite{Brunner:2007ur}. They can be represented in terms of $\bZ_d\times\bZ_d$-equivariant matrix factorizations $R$ of rank $d'$ which are parametrized by a choice of 
 $m\in\bZ_{d}$ and $n_0, ..., n_{d'-1}\in\bN_{>0}$ which sum up to $d=n_0 + ... + n_{d'-1}$. These RG  defects give rise to counital\footnote{Similarly, $\bP:=R^\dagger\otimes R$ are unital projection defects realizing the IR theory within the theory at the UV.} projection defects $P:={}^\dagger R\otimes R$ realizing the IR theory $X^{d'}/\bZ_{d'}$ within the UV theory $X^d/\bZ_d$ \cite{Klos:2019axh}.  They are represented by the $\bZ_d\times\bZ_d$-equivariant matrix factorizations
 \uaeq{P:P_1\;\tikz[baseline=0]{
			\begin{scope}
				\node at (0,2) {$\begin{pmatrix}Z^{n_1}& 0 & ... & 0 & -X^{n_0} \\ -X^{n_1} & Z^{n_2} &&&\\0&-X^{n_2}&\ddots&&\\\vdots&&\ddots&Z^{n_{d'-1}}&\\0&&&-X^{n_{d'-1}}&Z^{n_0}\end{pmatrix}$};
				\draw[arrow position = 1] (-3,.2) -- (3,.2);
				\draw[arrow position = 1] (3,0) -- (-3,0);
				\node at (0,-.3) {$\d_{P0}$};
			\end{scope}
		}\;P_0=S^{d'}\left(\begin{smallmatrix}
 	\{[r_0],[-r_0]\} \\
 	\{[r_1],[-r_1]\} \\
 	\{[r_2],[-r_2]\} \\
 	\vdots \\
 	\{[r_{d'-1}],[-r_{d'-1}]\}\end{smallmatrix}\right)
	}
of $Z^d-X^d$. Here, $S=\bC[X,Z]$, $\{\cdot,\cdot\}$ indicates the $\bZ_d\times\bZ_d$-charge of the respective $S$-module generator, and $[r_i]=[m+n_1+...+n_i]\in\bZ_{d}$. The charges of $P_1$ can easily be inferred from the ones of $P_0$. For more details see \cite{Klos:2019axh}. To simpify notation, we will consider the indices $i$ of $r_i$ and $n_i$ to be defined modulo $d'$. 

A counit $c:P\rightarrow I$ is given by
\[
	(c_0)_{ji} = -\frac{d'-1}{d'} \delta_{j,r_i},\qquad (c_1)_{kj} = -\frac{d'-1}{d'} \delta_{j,j_k} Z^{n_{j+1}-1-[k-r_j]}X^{[k-r_j]}\,.
\]
Here, $[k-r_j]$ denotes the representative of $(k-r_j)\text{ mod } d$ in $\{0,\ldots,d-1\}$, and $j_k$ the unique $j\in\bZ_{d'}$ that minimizes $[k-r_j]$.
Indeed, this counit is nothing but the evaluation map $P:={}^\dagger R\otimes R\rightarrow I$ which has been explicitely determined for the case at hand in \cite{Klos:2019axh} based on \cite{Carqueville:2012st,Carqueville:2012dk}.

Given the counit, a straightforward computation reveals that the complementary projection defect $\bP$ of $P$ is isomorphic to a direct sum 
\begin{equation}\label{eq:direct sum complementary projector}
	\bP = \bigoplus_{\substack{i\in\bZ_{d'}\\\text{with }n_i> 1}} \bP_i.
\end{equation}
of multiple unital projection defects ${\bar P}_i$, one for each $n_i>1$. The summands $\bP_i$ are given by the rank-$n_i$ matrix factorizations
 \uaeq{\bP_i:
 S^{n_i}\;\tikz[baseline=0]{
			\begin{scope}
				\node at (0,2) {$\begin{pmatrix}-X& Z & & & \\  & -X &Z&&\\ &&\ddots&\ddots&\\&&&-X&Z\\ Z^{d-n_i+1}&&&&-X^{d-n_i+1}\end{pmatrix}$};
				\draw[arrow position = 1] (-3,.2) -- (3,.2);
				\draw[arrow position = 1] (3,0) -- (-3,0);
				\node at (0,-.3) {$\d_{P0}$};
			\end{scope}
		}\;S^{n_i}\left(\begin{smallmatrix}
 	\{[r_{i-1}+1],[-r_{i-1}-1]\} \\
 	\{[r_{i-1}+2],[-r_{i-1}-2]\} \\
 	\{[r_{i-1}+3],[-r_{i-1}-3]\} \\
 	\vdots \\
 	\{[r_{i}-1],[-r_{i}+1]\} \\
 	\{[r_{i}],[-r_{i-1}]\}
 	\end{smallmatrix}\right)
	}
	It is easy to check that each $\bP_i$ is a projection defect with $\bP_i\otimes\bP_j\cong\delta_{i,j}\bP_i$, and that it is unital. In fact, it factorizes as $\bP_i\cong R_i^\dagger\otimes R_i$, where each $R_i$ is an RG defect from $X^d/\bZ_d$ to $X^{n_i}/\bZ_{n_i}$. Therefore, $\bP_i$ realizes the Landau-Ginzburg orbifold theory
$X^{n_i}/\bZ_{n_i}$ inside $X^d/\bZ_d$.

We obtain the following picture. The counital projector $P={}^\dagger R\otimes R$ associated to an RG defect from $X^d/\bZ_d\rightarrow X^{d'}/\bZ_{d'}$ specified by $(m,n_0,\ldots,n_{d'-1})$ projects the theory $X^d/\bZ_d$ in the UV to  the  IR theory $X^{d'}/\bZ_{d'}$. The respective complementary unital projection defect $\bP$ projects to the stack\footnote{Such stacks are sometimes also referred to disjoint unions \cite{Sharpe:2019ddn}, because sigma models whose target spaces are disjoint unions are particular examples.}
$$
\bigoplus_{\substack{i\in\bZ_{d'}\\\text{with }n_i>1}} X^{n_i}/ \bZ_{n_i}
$$ 
of Landau-Ginzburg orbifold theories.
\[\begin{tikzcd}
& X^d / \bZ_d \arrow{ld}[swap]{P} \arrow{rd}{\bP} & \\
X^{d'}/\bZ_{d'} & & \bigoplus\limits_{\substack{i\in\bZ_{d'}\\\text{with }n_i>1}} X^{n_i}/ \bZ_{n_i}
\end{tikzcd}\]
In fact, the mirror of the RG flow described by the defect $R$  is a flow from the unorbifolded Landau-Ginzburg model with superpotential $W=X^d$ to the one with superpotential $W'=X^{d'}$  triggered by a perturbation of the superpotential by lower order  terms. The parameters $n_i$ specify how the critical points of $W$ behave under the flow. More precisely, the $(d-1)$-times degenerate cricial point of $W=X^d$  breaks up into a $(d'-1)$-times degenerate critical point which is associated to  the IR theory, and $(n_i-1)$-times degenerate cricial points for each $n_i>1$. The latter run off to infinity and decouple from the theory. The complementary projection defect $\bP$ projects onto  all the decoupling parts of  the theory.

We will conclude this example by discussing the decomposition of the category of boundary conditions of the UV  theory with respect to the complementary pair $(P,\bP)$. 

Boundary conditions of a given theory are a special kind of defects, namely defects between the theory and the trivial theory. The trivial Landau-Ginzburg theory is the theory without any chiral superfields, and hence, B-type boundary conditions of the Landau-Ginzburg orbifold theory  $X^d/\bZ_d$ can be described by $\bZ_d$-equivariant matrix factorizations of $X^d$. It turns out that the category of these matrix factorizations is generated by the rank-$1$ linear matrix factorizations
\[B_{[c]}^1:\bC[X]\{[c+1]\}\;\tikz[baseline=0]{
		\begin{scope}
			\node at (0,.6) {$X^1$};
			\draw[arrow position = 1] (-1,.2) -- (1,.2);
			\draw[arrow position = 1] (1,0) -- (-1,0);
			\node at (0,-.3) {$X^{d-1}$};
		\end{scope}
	}\;\bC[X]\{[c]\}.
\]
Here $[c]\in\bZ_d$ determines the $\bZ_d$-action. All other $\bZ_d$-equivariant matrix factorizations of $X^d$ can be obtained from the $B_{[c]}^1$ via successive cone constructions.
For instance, any rank-$1$ $\bZ_d$-equivariant matrix factorization 
\[B_{[c]}^k:\bC[X]\{[c+k]\}\;\tikz[baseline=0]{
		\begin{scope}
			\node at (0,.6) {$X^k$};
			\draw[arrow position = 1] (-1,.2) -- (1,.2);
			\draw[arrow position = 1] (1,0) -- (-1,0);
			\node at (0,-.3) {$X^{d-k}$};
		\end{scope}
	}\;\bC[X]\{[c]\}
\]
of $X^d$
can be expressed via \eqref{eq:LG cones} as cone of a morphism $\phi=(-1,X^{d-k}):B^1_{[c]}[1]\rightarrow B^{k-1}_{[c+1]}$,
\[B_{[c]}^k\cong\bC[X]^2\begin{pmatrix}\{[c+k]\} \\ \{[c+1]\}\end{pmatrix}\;
		\tikz[baseline=0]{
			\node at (0,1) {$\begin{pmatrix}X^{k-1}& -1\\ &X\end{pmatrix}$};
			\draw[arrow position = 1] (-1.5,.2) -- (1.5,.2);
			\draw[arrow position = 1] (1.5,0) -- (-1.5,0);
			\node at (0,-.8) {$\begin{pmatrix}X^{d-k+1}&X^{d-k}\\  &X^{d-1}\end{pmatrix}$};
		}\;\;\bC[X]^2\begin{pmatrix}\{[c+1]\} \\ \{[c]\}\end{pmatrix}.
\]
Thus, by induction on $k$, any rank-$1$ matrix factorization $B_{[c]}^k$ can be obtained as successive cone of linear matrix factorizations $B_{[c]}^1$.
 
Employing methods of \cite{Brunner:2007ur,Brunner:2007qu}  to calculate fusion, we find that the action of $P$ on the linear boundary conditions of the UV theory $X^d/\bZ_d$ is given by
\begin{equation}\label{eq:linear boundary image}
	P\otimes B^1_{[c]} \cong\begin{cases}
		B^{n_{i+1}}_{[c]} &\quad\text{if }[c]=[r_i]\text{ for some }i\\
		0&\quad\text{otherwise}
	\end{cases}.
\end{equation}
Since $[r_i]\neq[r_j]$ for $i\neq j$, this implies that $d'$ linear boundary conditions have a non-trivial image under fusion with $P$. Since fusion commutes with the cone construction, the images $B^{n_{i+1}}_{[r_i]}$ generate the category of $P$-invariant boundary conditions. The latter corresponds to the category of 
boundary  condition of the $P$-projected theory, and is indeed
 isomorphic to the category of boundary conditions of the IR theory $X^{d'}/\bZ_{d'}$.

On the other hand there are $(d-d')$ linear boundary conditions $B^1_{[c]}$, $[c]\notin\{[r_i]\,|\,i\}$ annihilated by the fusion with $P$. They generate the $\bP$-invariant subcategory, i.e.~the category of boundary conditions of the  $\bP$-projected theory. Indeed, the summands $\bP_i$ of $\bP$ fuse with linear boundary conditions  according to 
\begin{equation}\label{eq:bP action on linear branes}
	\bP_i\otimes B^1_{[c]} \cong\begin{cases}
		B^1_{[c]} &\quad\text{if } [c]\in\{[r_{i-1}+1], [r_{i-1}+2],...,[r_i-1] \}\\
		B^{d-n_i+1}_{[r_i]} &\quad\text{if }[c]=[r_{i-1}] \\
		0&\quad\text{otherwise}
	\end{cases}.
\end{equation}

We explicitly  recover  the decomposition of the category of boundary conditions into the $P$-invariant and $\bP$-invariant subcategories (c.f.~the discussion around \eqref{eq:boundary composition}). The category of boundary conditions of $X^d/\bZ_d$ is  generated by the linear boundary conditions $B^1_{[c]}$. The ones with 
$[c]\notin \{[r_i]\,|\,i\}$ are annihilated by $P$ and hence are $\bP$-invariant. 
Moreover, for all $i$ with $n_i=1$ the linear boundary conditions $B^1_{[r_{i-1}]}$ are $P$-invariant:
 $P\otimes B^1_{[r_{i-1}]}\cong B^1_{[r_{i-1}]}$ and $\bP\otimes B^1_{[r_{i-1}]}\cong 0$.
The only linear boundary conditions which are neither $P$- nor $\bP$-invariant are the  
 $B^1_{[r_{i-1}]}$ with $n_i>1$.
 For these, \eqref{eq:linear boundary image} and \eqref{eq:bP action on linear branes} yield
\[
	P\otimes B^1_{[r_{i-1}]} \cong B^{n_i}_{[r_{i-1}]} \quad\text{and}\quad \bP \otimes B^1_{[r_{i-1}]} \cong B^{d-n_i+1}_{[r_{i}]}\,.
\]
Hence, they can be represented as cones of morphisms
$$
\bP\otimes B^1_{[r_{i-1}]}[-1]\cong B^{d-n_i+1}_{[r_{i}]}[-1]\longrightarrow P\otimes B^1_{[r_{i-1}]}\cong B^{n_i}_{[r_{i-1}]}
$$ 
between objects of the $\bP$-invariant and the $P$-invariant subcategories, c.f.~\eqref{eq:boundary composition}. Explicitly,
\[
	B^1_{[r_{i-1}]} \cong \bC[X]^2\begin{pmatrix}\{[r_i]\} \\ \{[r_{i-1}+1]\}\end{pmatrix}\;
		\tikz[baseline=0]{
			\node at (0,1) {$\begin{pmatrix}X^{n_i}& X\\ &X^{d-n_i+1}\end{pmatrix}$};
			\draw[arrow position = 1] (-1.5,.2) -- (1.5,.2);
			\draw[arrow position = 1] (1.5,0) -- (-1.5,0);
			\node at (0,-.8) {$\begin{pmatrix}X^{d-n_i}&-1\\  &X^{n_i-1}\end{pmatrix}$};
		}\;\;\bC[X]^2\begin{pmatrix}\{[r_{i-1}]\} \\ \{[r_i]\}\end{pmatrix}\,.
\]
The decomposition on the level of the generators determines the decomposition of the entire category of boundary conditions into $P$- and $\bP$-invariant subcategories.  

\section{Conclusions}

In this note we have shown that in theories whose defect categories are tensor triangulated, projection defects 
always come in complementary pairs $(P,\bP)$. These have the following properties. $P$ is counital and $\bP$ is unital, and the identity defect is isomorphic to a cone of a morphism $s:\bP[-1]\rightarrow P$.

(Co)unital projection defects always split, i.e.~there are RG-type defects $\bar{R}$ ($R$) such that $\bP\cong\bar{R}^\dagger\otimes\bar{R}$ ($P\cong{}^\dagger R\otimes R$), and the projection defects realize projected theories in the given host theory \cite{Klos:2019axh}. The fact that in the triangulated setup, 
projection defects always come in complementary  pairs 
means that whenever there is a projection defect in this context, the host theory
 decomposes into the projected theory associated to $P$ and the complementary projected theory associated to $\bP$.

In the explicit example we considered, the counital projection defect $P$ splits as $P\cong{}^\dagger R\otimes R$, where $R$ is the RG defect associated to a renormalization group flow between the Landau-Ginburg  orbifolds $X^d/\bZ_d$ in the UV and $X^{d'}/\bZ_{d'}$ in the IR ($d>d'$). Hence, $P$ realizes the IR theory inside the UV.
The complementary  projection defect $\bP$ in this case collects all the parts  of the theory which decouple during the RG flow; and the UV theory deomposes into two parts, the IR theory on the one hand and the decoupling parts on the other.

An interesting open question is the physical significance  of  the fact that  the identity defect in the host theory is  isomorphic to the cone of some morphism $s:\bP[-1]\rightarrow P$.  Obviously this means that $P\oplus\bP$ deforms to the identity defect. If $P\oplus\bP$ was itself a 
(co)unital projection defect, with a well-defined projected theory  associated to it, then this implies that the latter can be deformed (or perturbed) to the host theory. 
However, in general $P\oplus\bP$ is neither unital nor counital. Hence, it is not clear to what extent $P\oplus\bP$ describes an honest topological quantum field theory. It would be interesting to shed some light on the physical significance of the sum $P\oplus\bP$ and the cone condition.

\section*{Acknowledgements}

FK is thankful to Friedrich-Naumann-Stiftung for supporting this project. DR is supported by the Heidelberg Institute for Theoretical Studies. DR also thanks the MSRI in Berkeley for its hospitality, where part of this work was done. (Research at MSRI is partly supported by the NSF under Grant No. DMS-1440140.)

	\bibliographystyle{JHEP}
	\bibliography{references}
	
\end{document}